\documentclass[aps,prl,reprint,showpacs,byrevtex]{revtex4-2}
\let\oldhat\hat
\renewcommand{\hat}[1]{\oldhat{\mathbf{#1}}}
\usepackage{titlesec}
\usepackage{array}

\usepackage{times,mathptm}
\usepackage[dvips]{graphicx,color}
\usepackage{amsmath}
\usepackage{xcolor}

\begin{document}
\title{Absence of ferromagnetic instability and weak spin-orbit coupling effect in AV$_3$Sb$_5$ (A = Cs, Rb, and K)}
\author{Chongze Wang$^{1*}$, Shichang Yao$^{1}$, Shuyuan Liu$^2$, Bing Wang$^{1}$, Liangliang Liu$^{3,4}$, Yu Jia$^{3,4}$, and Jun-Hyung Cho$^{1,2*}$}
\affiliation{$^1$Joint Center for Theoretical Physics, School of Physics and Electronics, Henan University, Kaifeng 475004, People's Republic of China \\
$^2$Department of Physics, Hanyang University, 222 Wangsimni-ro, Seongdong-Ku, Seoul 04763, Republic of Korea \\
$^3$Key Laboratory for Special Functional Materials of the Ministry of Education, Henan University, Kaifeng 475004, People's Republic of China \\
$^4$Institute of Quantum Materials and Physics, Henan Academy of Sciences, Zhengzhou 450046, China}
\date{\today}

\begin{abstract}
A family of V-based kagome metals AV$_3$Sb$_5$ (A = Cs, Rb, K) presents an intriguing platform for exploring the interplay of time-reversal symmetry breaking, nontrivial topological bands, and electron correlations, resulting in a range of exotic quantum states, including the anomalous Hall effect, unconventional charge density waves, and superconductivity. These features prompt critical questions regarding the roles of magnetism and spin-orbit coupling (SOC) in these systems. Our density functional theory (DFT) calculations demonstrate a notable sensitivity of the magnetic properties to the choice of $k$-point mesh used in Brillouin zone integrations. Specifically, we find that using a dense $k$-point mesh yields a nonmagnetic pristine phase characterized by paramagnetic susceptibility, consistent with the recently observed Pauli paramagnetic behavior in single crystalline samples at high temperatures. In contrast, a coarser $k$-point mesh significantly increases the density of states at the Fermi level, inducing a ferromagnetic instability that satisfies the Stoner criterion. Moreover, our results show that the effect of SOC on both the geometric and electronic structures is minimal, with only a slight gap opening at the Dirac points, indicating a weak SOC influence in these materials. Importantly, our DFT band structure calculations closely align with angle-resolved photoemission spectroscopy data, reinforcing the notion of weak electron correlations in these kagome metals. This refined understanding challenges recent theoretical assertions that the interplay of magnetism, SOC, and electron correlations is essential for determining the nature of charge density waves in AV$_3$Sb$_5$.
\end{abstract}
\maketitle

\section{I. INTRODUCTION}

The kagome metals AV$_3$Sb$_5$ (A = Cs, Rb, K) have attracted significant attention due to their intriguing electronic properties~\cite{Ortiz-PRM2019, Ortiz-PRL2020, AV3Sb5_rev_NatPhy, AV3Sb5_rev_JAP}, including the emergence of charge density waves (CDWs) and superconductivity at low temperatures~\cite{KV3Sb5-chrialCDW-Nat.Mat2021, KV3Sb5_SC_Z2_PRM2021, RbV3Sb5-SC-CPL2021, CsV3Sb5-CDW_SC-NC2021, CsV3Sb5-SC_CDW-PRL2021}. These materials feature a unique lattice geometry that has been proposed to give rise to exotic quantum states, such as topological phases and potential time-reversal symmetry-breaking (TRSB) states~\cite{Binghai_PRL2021, Jiangping_SB2021}. A key point of contention in the study of AV$_3$Sb$_5$ is the presence of local magnetic moments on V atoms. In the pioneering work of Ortiz \textit{et al.}~\cite{Ortiz-PRM2019} in 2019, magnetization measurements on KV$_3$Sb$_5$ exhibited Curie-Weiss behavior at high temperatures, with an effective magnetic moment of approximately 0.22 ${\mu}_B$ per V atom. However, neutron scattering measurements conducted by Ortiz \textit{et al.}~\cite{Ortiz-PRM2019} did not detect signatures of long-range or short-range magnetic ordering below the magnetization anomaly. Subsequent muon spin spectroscopy~\cite{Kenney-JPCM-2021} demonstrated the absence of local magnetic moments in KV$_3$Sb$_5$, revealing no evidence of conventional local electronic moments. Moreover, recent experimental studies~\cite{Ortiz-PRL2020} on single-crystalline samples have demonstrated that the high-temperature magnetic susceptibility from magnetization measurements under an applied magnetic field is Pauli-like, suggesting that the Curie-Weiss-like behavior observed in earlier powder samples likely originated from local impurity spins rather than intrinsic V moments.

Several theoretical models have proposed mechanisms for the formation of TRSB states~\cite{KV3Sb5-chrialCDW-Nat.Mat2021,Jiangping_SB2021,Nature_Chiraltrans} in AV$_3$Sb$_5$, such as loop currents~\cite{Jiangping_SB2021,Nature_Chiraltrans} and magnetism~\cite{Hasan-PRL-2023}. Specifically, Hasan \textit{et al.}~\cite{Hasan-PRL-2023} employed various first-principles approaches, including density functional theory (DFT), DFT with spin-orbit coupling (SOC), and DFT combined with dynamical mean-field theory (DMFT) and SOC, to argue for the existence of local magnetic moments on V atoms and their impact on electronic and vibrational properties. They proposed that the interplay between magnetism, SOC, and electron correlations significantly influences the electronic structure, phonon dispersion, and electron-phonon coupling, offering explanations for the nature of charge density waves (CDWs) and superconductivity in AV$_3$Sb$_5$. However, their report of pronounced SOC effects, including substantial exchange splitting, is unusual given that SOC effects in 3$d$ electron systems are generally expected to be small. Additionally, their DFT + DMFT + SOC band structure calculations do not align with angle-resolved photoemission spectroscopy (ARPES) data around van Hove singularities (VHSs) and Dirac points near the Fermi level ($E_F$). In contrast, the ARPES-measured band dispersions of AV$_3$Sb$_5$ closely matches those predicted by standard DFT calculations~\cite{arpes1,arpes2,135_DNL}, indicating minimal band renormalization due to correlation effects. To reconcile these discrepancies and enhance our understanding of the fundamental mechanisms underlying the unique electronic properties of these kagome metals, further investigation into the potential role of magnetism and the effects of SOC in AV$_3$Sb$_5$ is necessary.

In the present study, we perform first-principles DFT and DFT + SOC calculations to explore the ferromagnetic (FM) instability in AV$_3$Sb$_5$ and evaluate the impact of SOC on its electronic band structure. Our results indicate that the ground state of the pristine phase of AV$_3$Sb$_5$ is nonmagnetic (NM), contrasting with a recent theoretical report~\cite{Hasan-PRL-2023} suggesting a FM state. We attribute this discrepancy to the use of an insufficiently dense $k$-point mesh in Brillouin zone integrations, which can artificially increase the density of states (DOS) at $E_F$, thereby fulfilling the Stoner criterion for ferromagnetism. Additionally, our analysis shows that the inclusion of SOC induces minimal changes in the geometry and band structure of AV$_3$Sb$_5$, primarily resulting in a slight gap opening at the Dirac points. Thus, our findings, together with the close alignment of our DFT-derived band structure with ARPES data~\cite{arpes1,arpes2,135_DNL}, provide a revised perspective on recent theoretical assertions~\cite{Hasan-PRL-2023} that the interplay of magnetism, SOC, and electron correlations is crucial for determining the nature of CDWs in AV$_3$Sb$_5$.

\begin{figure}[h!t]
\includegraphics[width=8.5cm]{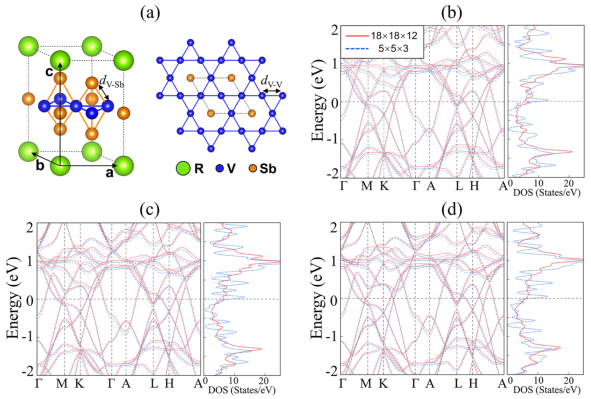}
\caption{(a) Optimized structure of the pristine phase of CsV$_3$Sb$_5$ and the corresponding DFT band structures and DOS for (b) CsV$_3$Sb$_5$, (c) RbV$_3$Sb$_5$, and (d) KV$_3$Sb$_5$. Solid and dashed lines represent the NM results obtained using 18${\times}$18${\times}$12 and 5${\times}$5${\times}$3 $k$-point meshes, respectively.}
\label{figure:1}
\end{figure}

\section{II. CALCULATIONAL METHODS}

We conducted first-principles calculations using DFT ad DFT + SOC as implemented in the Vienna ab initio Simulation Package (VASP)~\cite{vasp1,vasp2}. The core potential was described using the projector augmented wave (PAW) method~\cite{paw}, with the following valence electron configurations: Cs (5$s^2$5$p^6$6$s^1$), Rb (4$s^2$4$p^6$5$s^1$), K (3$s^2$3$p^6$4$s^1$), V (3$p^6$3$d^4$4$s^1$), and Sb (5$s^2$5$p^3$). We included SOC for all valence orbitals in AV$_3$Sb$_5$, including the V 3$d$ and Sb 5$p$ states, ensuring a thorough treatment of relativistic effects. For the exchange-correlation interactions, we employed the Perdew-Burke-Ernzerhof (PBE) generalized gradient approximation functional~\cite{pbe}. The van der Waals interactions were included using the DFT-D3 scheme~\cite{DFT-D3-zero}. A plane wave basis set with a kinetic energy cutoff of 500 eV was used, and $k$-space integration was performed with an 18${\times}$18${\times}$12 $k$-mesh. Atomic positions were relaxed until the residual force components on each atom were less than 0.001 eV/{\AA}. These computational parameters have been successfully applied in our previous study of CDWs in CsV$_3$Sb$_5$~\cite{chongze_PRB2022,chongze_PRM2022,chongze_PRB2023} and RbV$_3$Sb$_5$~\cite{chongze_PRB2022, sc_PRM2023}.

\section{III. RESULTS AND DISCUSSION}

\begin{table*}[ht]
\caption{Calculated lattice parameters ($a$, $b$, and $c$) and bond lengths ($d_{\rm V-V}$ and $d_{\rm V-Sb}$) for AV$_3$Sb$_5$ in both NM and FM states, obtained using different $k$-meshes.}
\begin{ruledtabular}
\begin{tabular}{lccccccc}
 & $k$-mesh & State & SOC & $a$ = $b$ ({\AA}) & $c$ ({\AA}) & $d_{\rm V-V}$ ({\AA}) & $d_{\rm V-Sb}$ ({\AA}) \\ \hline
CsV$_3$Sb$_5$ & 18${\times}$18${\times}$12 & NM  & No & 5.438 &  9.354 & 2.719 & 2.754 \\
              & 18${\times}$18${\times}$12 & NM  & Yes & 5.440 &  9.335 & 2.720 & 2.754 \\
              & 5${\times}$5${\times}$3    & NM  & No & 5.430 &  9.406 & 2.715 & 2.759 \\
              & 5${\times}$5${\times}$3    & FM  & No & 5.433 &  9.396 & 2.716 & 2.758 \\
RbV$_3$Sb$_5$ & 18${\times}$18${\times}$12 & NM  & No & 5.423 &  9.092 & 2.712 & 2.759 \\
              & 18${\times}$18${\times}$12 & NM  & Yes & 5.426 &  9.084 & 2.713 & 2.759 \\
              & 5${\times}$5${\times}$3    & NM  & No & 5.419 &  9.128 & 2.709 & 2.763 \\
              & 5${\times}$5${\times}$3    & FM  & No & 5.419 &  9.128 & 2.710 & 2.763 \\
KV$_3$Sb$_5$  & 18${\times}$18${\times}$12 & NM  & No & 5.412 &  8.907 & 2.706 & 2.764 \\
              & 18${\times}$18${\times}$12 & NM  & Yes & 5.414 &  8.901 & 2.707 & 2.764 \\
              & 5${\times}$5${\times}$3    & NM  & No & 5.407 &  8.943 & 2.704 & 2.768 \\
              & 5${\times}$5${\times}$3    & FM  & No & 5.408 &  8.943 & 2.704 & 2.768 \\
\end{tabular}
\end{ruledtabular}
\end{table*}

We first investigate the FM instability for the pristine phase of AV$_3$Sb$_5$ using spin-polarized DFT calculations with an 18${\times}$18${\times}$12 $k$-mesh, excluding SOC. Our results indicate that the pristine phase is definitively NM, as any initial configuration of magnetic moments quickly converges to a NM solution. Figure 1(a) shows the optimized structure of CsV$_3$Sb$_5$, which crystallizes in the hexagonal space group $P6/mmm$ (No. 191). The structure consists of a V$_3$Sb kagome layer with a triangular Sb sublattice centered on each V hexagon, Sb honeycomb layers situated above and below the V$_3$Sb kagome layer, and a Cs triangular layer. The calculated lattice parameters [$a$, $b$, and $c$, as shown in Fig. 1(a)] and bond lengths [V$-$V and V$-$Sb, denoted as $d_{\rm V-V}$ and $d_{\rm V-Sb}$ in Fig. 1(a)] for AV$_3$Sb$_5$ are summarized in Table I. Due to variations in the lattice parameters $a$, $b$, and $c$ among CsV$_3$Sb$_5$, RbV$_3$Sb$_5$, and KV$_3$Sb$5$, their bond lengths differ slightly, with $d_{\rm V-V}$ ($d_{\rm V-Sb}$) values of 2.719 (2.754) {\AA}, 2.712 (2.759) {\AA}, and 2.706 (2.764) {\AA}, respectively. Hereafter, CsV$_3$Sb$_5$, RbV$_3$Sb$_5$, and KV$_3$Sb$_5$ will be referred to as CVS, RVS, and KVS, respectively. The electronic band structures of the pristine phases of CVS, RVS, and KVS, along with their corresponding DOS, are displayed in Figs. 1(b), 1(c), and 1(d), respectively. The calculated DOS at $E_F$ is 6.58, 6.42, and 6.16 states/eV per unit cell for CVS, RVS, and KVS, respectively.

\begin{figure*}[h!t]
\includegraphics[width=17cm]{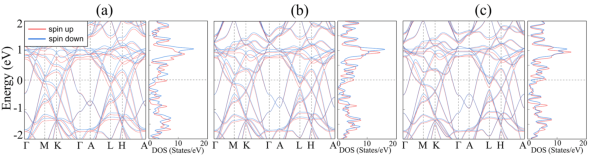}
\caption{Calculated FM band structures of (a) CsV$_3$Sb$_5$, (b) RbV$_3$Sb$_5$, and (c) KV$_3$Sb$_5$ using the 5${\times}$5${\times}$3 $k$-mesh, along with the corresponding DOS for spin-up and spin-down electrons. The exchange splitting ($E_{\rm up}-E_{\rm down}$) between spin-up and spin-down states is estimated by calculating the difference between their Kohn-Sham eigenvalues: i.e., $\int_{-\infty}^{E_{\rm up}} {D}_{\rm up}(E) dE = N_e$, and $\int_{-\infty}^{E_{\rm down}} {D}_{\rm down} (E) dE = N_e$, where $N_e$ is the occupation number of spin-up and spin-down electrons in the NM state. Here, $D_{\rm up}$ and $D_{\rm down}$ represent the DOS for the spin-up and spin-down bands, respectively.}
\label{figure:2}
\end{figure*}

Hasan \textit{et al.}~\cite{Hasan-PRL-2023} previously reported, based on DFT calculations with a 5${\times}$5${\times}$3 $k$-point mesh, that the FM state can be stabilized with local magnetic moments on the V atoms in CVS, RVS, and KVS (see Supplemental Table S1 in Ref.~\cite{Hasan-PRL-2023}). To verify these findings, we performed spin-polarized DFT calculations using the same 5${\times}$5${\times}$3 $k$-mesh. Our results show that the FM state is energetically favored over the NM state by ${\Delta}E$ = 9.54, 3.48, and 1.45 meV per unit cell for CVS, RVS, and KVS, respectively. The magnetic moment of the V atom, calculated using the PAW method, is 0.15, 0.10, and 0.08 ${\mu}_B$ for CVS, RVS, and KVS, respectively. As shown in Table I, the NM and FM structures of CVS, RVS, and KVS obtained with the 5${\times}$5${\times}$3 $k$-mesh exhibit minimal changes in lattice parameters, $d_{\rm V-V}$, and $d_{\rm V-Sb}$ compared to those obtained with the 18${\times}$18${\times}$12 $k$-mesh. The dashed lines in Figs. 1(b), 1(c), and 1(d) represent the NM band structures of CVS, RVS, and KVS obtained with the 5${\times}$5${\times}$3 $k$-mesh. The DOS values at $E_F$ for CVS, RVS, and KVS are 10.97, 8.66, and 7.85 states/eV per unit cell, respectively, which are higher than those obtained with the 18${\times}$18${\times}$12 $k$-mesh. The increased DOS at $E_F$ with the 5${\times}$5${\times}$3 $k$-mesh results in Stoner criterion $I \cdot D(E_F)$ exceeding 1, with values of 1.82, 1.44, and 1.21 for CVS, RVS, and KVS, respectively. Here, $D(E_F)$ is given in units of states/eV per spin per unit cell, and the Stoner parameter $I$ is estimated to be 0.33, 0.33, and 0.31 eV for CVS, RVS, and KVS, respectively. These values are calculated by dividing the exchange splitting values of 0.15, 0.11, and 0.08 eV [see Figs. 2(a), 2(b), and 2(c)] by the magnetic moments of 0.46, 0.32, and 0.26 ${\mu}_B$ per unit cell. This demonstrates that using a 5${\times}$5${\times}$3 $k$-mesh can induce FM instability via the Stoner mechanism, along with the exchange splitting of V 3$d$ orbitals. It is important to note that the differences between the 5${\times}$5${\times}$3 and 18${\times}$18${\times}$12 $k$-meshes underscore the significant impact of Brillouin zone sampling on magnetic properties. A coarser mesh may inadequately sample the Brillouin zone, leading to inaccurate electronic structures and potentially incorrect predictions of the ground state, which can result in an erroneous determination of the FM state. In contrast, a denser mesh can reveal a more stable NM state, correcting the artifacts that favored the FM state with the coarser mesh.

Using the fixed-spin-moment method~\cite{Williams84}, we further investigate the FM instability predicted with the 5${\times}$5${\times}$3 $k$-mesh and the NM ground state determined with the 18${\times}$18${\times}$12 $k$-mesh. Figure 3 illustrates the variation of the total energy as a function of magnetization ($M$) for CVS, RVS, and KVS. The results with the denser 18${\times}$18${\times}$12 $k$-mesh show that the total energy decreases monotonically as $M$ approaches zero, indicating a NM state. In contrast, the coarser 5${\times}$5${\times}$3 $k$-mesh stabilizes a FM state with $M$ of approximately 0.48, 0.31, and 0.27 ${\mu}_B$ for CVS, RVS, and KVS, respectively, as indicated by the arrows in Fig. 3. These differences between the 18${\times}$18${\times}$12 and 5${\times}$5${\times}$3 $k$-meshes arise from the competition between kinetic energy and exchange interaction energy as a function of $M$. In the 5${\times}$5${\times}$3 $k$-mesh, increasing $M$ incurs a kinetic energy cost due to electron redistribution between spin states. However, the gain in exchange energy, which favors spin alignment, stabilizes the FM state. Thus, the FM stabilization predicted with the 5${\times}$5${\times}$3 $k$-mesh reflects the balance between kinetic and exchange energies. On the other hand, this balance is more accurately captured with the denser 18${\times}$18${\times}$12 $k$-mesh, which stabilizes the NM state by imposing a higher kinetic energy cost for magnetization. From the total energy versus $M$ curves obtained using the 18${\times}$18${\times}$12 $k$-mesh, we estimate the magnetic susceptibility ${\chi}$, defined as ${\chi} = \left(\frac{{\partial}^2E}{{\partial}M^2}\right)^{-1}$~\cite{Mazin}, to be 8.48, 8.39, and 6.51 ${\times}$10$^{-4}$ emu/mol for CVS, RVS, KVS, respectively. This feature is consistent with the recently observed Pauli paramagnetic behavior in single crystalline samples of CVS at high temperatures above 100 K~\cite{Ortiz-PRL2020}, where ${\chi}$ was measured to be approximately 3.5${\times}$10$^{-4}$ emu/mol for CVS. The somewhat overestimated ${\chi}$ values in our zero-temperature DFT calculations is likely due to the neglect of spin fluctuations in the mean-field approximation of DFT~\cite{Moriya, Larson}.

\begin{figure}[ht]
\centering{\includegraphics[width=7.5 cm]{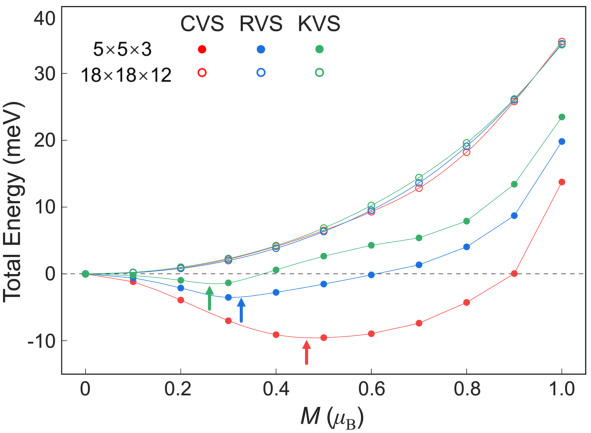}}
\caption{Total energy variation as a function of the magnetic moment $M$ per unit cell for CVS, RVS, and KVS, calculated using the 5${\times}$5${\times}$3 and 18${\times}$18${\times}$12 ${k}$-meshes. Each NM configuration with $M$ = 0 defines the energy zero. The arrows in the 5${\times}$5${\times}$3 results indicate the positions of the minimum energy.}
\label{figure:1}
\end{figure}

Next, we examine the effect of SOC on the geometry and band structure of AV$_3$Sb$_5$. As shown in Table I, including SOC with the 18${\times}$18${\times}$12 $k$-mesh results in only minor changes to the lattice parameters and V–V and V–Sb bond lengths, with variations mostly less than 0.01 {\AA}, indicating that SOC has a negligible effect on the geometry of AV$_3$Sb$_5$. The calculated DFT and DFT + SOC band structures of CVS are compared in Fig. 4(a), along with the ARPES data~\cite{arpes1}. The corresponding band structures of RVS and KVS are provided in Supplemental Fig. S1~\cite{SM}. We find that SOC leads to only minimal deviations from the DFT band structure, except for a gap opening of less than ${\sim}$20 meV at the Dirac points, which aligns well with previous DFT and DFT + SOC calculations~\cite{arpes2,135_DNL}. This suggests that SOC has a limited impact on the band structures of AV$_3$Sb$_5$. In contrast, a recent DFT + SOC calculation of Hasan \textit{et al.}~\cite{Hasan-PRL-2023} reported that the inclusion of SOC significantly alters their DFT band structure [see Figs. 4(b) and 4(c)]. For instance, the DFT + SOC band structure of CVS shows not only a large exchange splitting with an enhanced spin moment of 0.776 ${\mu}_B$ per V atom (refer to Table S3 in Ref.~\cite{Hasan-PRL-2023}), but also substantial upward shifts in the bands, with some flatbands above $E_F$ being shifted by more than 0.5 eV compared to their DFT band structure [see the dashed boxes in Figs. 4(b) and 4(c)]. These significant SOC effects in V-based compounds are surprising, given the generally weak SOC strength associated with 3$d$ electrons. We note that the SOC-induced gaps at Dirac points or nodal lines in 3$d$ transition metal oxides such as TiO$_2$~\cite{TiO2_NC2019} and FeO~\cite{FeO_JCP1985} are in the range of 1$-$10 meV, while those in iron-based superconductors~\cite{FeSe_NP2015} are around 10$-$20 meV.

\begin{figure}[h!t]
\includegraphics[width=8.5cm]{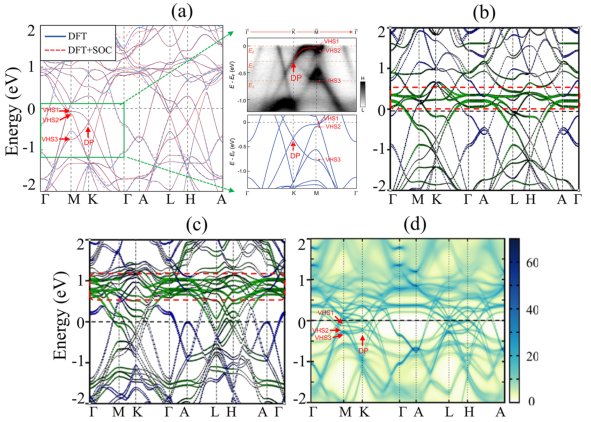}
\caption{(a) Comparison of the DFT and DFT + SOC band structures of CVS with ARPES data~\cite{arpes1}. The inset includes the previously reported~\cite{arpes1} DFT band structure. For comparison, the DFT, DFT + SOC, and DFT + DMFT + SOC band structures of CVS reported by Hasan \textit{et al.}~\cite{Hasan-PRL-2023} are shown in (b), (c), and (d), respectively.}
\label{figure:4}
\end{figure}

It has been known that the AV$_3$Sb$_5$ family materials are weakly correlated metals in their pristine and CDW phases~\cite{DMFT1,DMFT2}. However, as shown in Fig. 4(d), Hasan \textit{et al.}'s DFT + DMFT + SOC calculations for the pristine CVS revealed significant band modifications due to dynamical electronic correlations. These modifications led to a reduction in the local spin moment of V to approximately 0.28 ${\mu}_B$, compared to a larger value of 0.776 ${\mu}_B$ predicted by their DFT + SOC calculations~\cite{Hasan-PRL-2023}. Although Hasan \textit{et al.} claimed that the DFT + DMFT + SOC value is closer to the experimental value of 0.22 ${\mu}_B$ reported in Ref.~\cite{Ortiz-PRM2019}, this experimental data has been questioned. The anomalies observed in both magnetization and heat capacity measurements of KV$_3$Sb$_5$ at 80 K (below which the local moment is largely quenched) was suggested to be attributed to impurity spins on the V sites~\cite{Ortiz-PRM2019}. Muon spin rotation experiments~\cite{Kenney-JPCM-2021} have indicated the absence of a measurable local spin moment on V atoms. Additionally, experiments on single crystals~\cite{Ortiz-PRL2020} showed that the high-temperature magnetic susceptibility exhibits Pauli-like behavior, consistent with Pauli paramagnetism. Therefore, the local moment initially reported in powder samples~\cite{Ortiz-PRM2019} is unlikely to stem from the intrinsic magnetic moment of the V atom but rather from impurity spins. This raises doubts about the reliability of Hasan \textit{et al.}'s DMFT results, tough their predicted magnetic moment of V apparently agrees with experimental data~\cite{Ortiz-PRM2019}.

As depicted in Fig. 4(a), our DFT band structure for CVS reveals three VHSs (designated as VHS1, VHS2, and VHS3) located at the $M$ point below $E_F$, along with a Dirac point at the $K$ point, approximately $-$0.3 eV below $E_F$. Notably, the positions of these VHSs and Dirac point show good agreement with ARPES measurements~\cite{arpes1,arpes2,135_DNL} [see Fig. 4(a)]. In contrast, the DFT + DMFT + SOC band structure presented by Hasan \textit{et al.} [Fig. 4(d)] does not accurately reproduce the ARPES-observed band dispersions, particularly the positions of VHSs and the band multiplicity around the Dirac point. This misalignment with experimental observations raises concerns about the robustness of their DFT + DMFT + SOC approach, especially regarding local magnetic moments on V atoms. These inconsistencies suggest that Hasan \textit{et al.}'s methods~\cite{Hasan-PRL-2023} may not fully capture the complex electronic structure of CVS, calling into question the reliability of their findings. In contrast, other DMFT calculations~\cite{DMFT1,DMFT2} indicate that the computed DFT band structure and DFT + DMFT quasiparticle dispersion are very similar near $E_F$, suggesting weak electronic correlations in AV$_3$Sb$_5$.

\section{IV. SUMMARY}

Our DFT and DFT + SOC calculations demonstrated that the pristine phase of the kagome metals AV$_3$Sb$_5$ possesses a NM ground state with the absence of local magnetic moments on V atoms, supported by experimental evidence from magnetization measurements~\cite{Ortiz-PRM2019,Ortiz-PRL2020}, elastic neutron scattering~\cite{Ortiz-PRM2019}, and muon spin rotation spectroscopy~\cite{Kenney-JPCM-2021}. We revealed that the previously reported~\cite{Hasan-PRL-2023} FM state in AV$_3$Sb$_5$ arises from the use of an insufficient $k$-point mesh for Brillouin zone integrations, which artificially increases the DOS at $E_F$, leading to a false satisfaction of the Stoner criterion for ferromagnetism. Additionally, we found that while SOC opens a topological gap at the Dirac points in AV$3$Sb$_5$, its influence on the geometry and electronic structure remains minimal, suggesting a weak SOC effect. Our findings, along with the close alignment of our DFT-derived band structure with ARPES data~\cite{arpes1,arpes2,135_DNL}, do not support recent theoretical conclusions~\cite{Hasan-PRL-2023} regarding the significant roles of magnetism, SOC, and electron correlations in AV$_3$Sb$_5$.

\vspace{0.4cm}

\noindent {\bf Acknowledgements.}
This work was supported by the National Research Foundation of Korea (NRF) grant funded by the Korean Government (Grant No. RS202300218998), the National Natural Science Foundation of China (Grants No. 12104130 and No. 12074099), and the Natural Science Foundation of Henan Province (Grants No. 242300421162 and No. 242300421213).



\noindent $^{*}$ Corresponding authors: czwang29@gmail.com, cho@henu.edu.cn

\end{document}